\documentclass[fleqn,10pt]{wlscirep}
\usepackage[utf8]{inputenc}
\usepackage[T1]{fontenc}
\usepackage{multirow} 
\usepackage{soul}
\usepackage{caption}
\usepackage{graphicx}
\title{Direct Evidence of a Highest Wave-Driven Energetic Electron Flux at the Earth's Magnetopause}

\author[1]{Shubhangi Lagad}
\author[1,*]{Amar Kakad}
\author[1]{Bharati Kakad}
\affil[1]{Indian Institute of Geomagnetism, New Panvel, Navi Mumbai, 410218, India}

\affil[*]{amar.kakad@iigm.res.in}

\begin{abstract}
Spacecraft observations of high-energy electron flux enhancement up to 125 keV at Earth's magnetopause are typically linked to the magnetic reconnection. Here, we report the first ever observation of prolonged electron flux enhancement reaching very high energy up to 650 keV at magnetopause near the subsolar point, notably without reconnection signatures. The observation reveals that the high-energy electron flux enhancement near the magnetopause is associated with the simultaneous occurrence of electron cyclotron wave harmonics, whistler waves, and Langmuir waves. Theoretical modeling confirms the wave generation by electron temperature anisotropy and the electron loss cone distribution. This direct, in-situ observation of high-energy electron flux linked to cascaded wave-particle interactions will help us advance our understanding about the tiniest electron-scale intricacies affecting near-Earth space weather.
\end{abstract}
\begin{document} 
\flushbottom
\maketitle
%
%
\thispagestyle{empty}


\section*{Introduction}

The Earth's magnetosphere is an exceptional, naturally occurring plasma laboratory that can be used to study basic plasma processes. The complex interaction between the solar wind and the geomagnetic field, via magnetic reconnection, facilitates the transfer of solar wind mass, momentum, and energy into the magnetosphere \cite{frey2003continuous, hesse2020magnetic}. Magnetic reconnection is fundamental in driving various plasma processes \cite{hwang2015magnetopause} on the day and nightside of the magnetosphere, impacting the near-Earth space weather. Among these processes, the electron acceleration near the reconnection region has received much attention since the launch of the Magnetospheric Multiscale (MMS) mission. While reconnection at the magnetopause rarely shows strong electron acceleration \cite{tang2013themis}, magnetotail observations reveal electron acceleration up to 200 - 300 keV \cite{oieroset2002evidence,wang2010observations}. Recent observations demonstrate that the acceleration in the magnetotail can significantly exceed previous estimates, leading to electron populations with energies up to several MeV \cite{zhang2024relativistic}.
Spacecraft observations show that the whistler waves, electron cyclotron waves and their harmonics (ECH), and waves around electron plasma frequency $f_{pe}$, play a key role in local electron heating, particularly during reconnection as these are high-frequency and localized waves \cite{khotyaintsev2019collisionless}. During magnetotail reconnection, whistler waves are associated with perpendicular heating of few to tens of keV electrons leading to a temperature anisotropy that promotes wave growth \cite{wei2007cluster}. Farrell et al.\cite{farrell2002dominance, farrell2003role} reported waves around electron plasma frequency during magnetotail reconnection. Whistler waves are often related to the dayside magnetic reconnection \cite{jaynes2016energetic, fu2019evidence, chepuri2022statistical, contel2016whistler, wilder2016observations, zhou2018magnetospheric}. Langmuir and ECH are also reported during magnetic reconnection \cite{farrell2002dominance, zhou2016observation}. There are reports about the simultaneous observations of whistler and Langmuir waves \cite{li2018local}, and whistler and ECH waves \cite{yu2021simultaneous} during magnetic reconnection.

The solar wind varies on the timescale from less than seconds to hours, causing the magnetopause, where the solar wind interacts with the geomagnetic field, is always in a state of dynamic behaviour. The magnetopause’s response to such varying input is rapid and abrupt leading to intensification of plasma waves and high energy fluxes. Jaynes et al.\cite{jaynes2016energetic} reported electron flux enhancement up to 100 keV and Fu et al.\cite{fu2019evidence} reported up to 4 keV during magnetic reconnection at the magnetopause. Chepuri et al.\cite{chepuri2022statistical} reported flux enhancements up to 100 keV from 2016 - 2020 magnetopause crossing events, many of them showing reconnection signatures.

The wave dynamical complexity further increases with intense solar eruptions that cause severe geomagnetic activities. There are several reports about the intensification of the electron waves \cite{peng2020variation, meredith2009survey, ni2017statistical, liu2020global, xie2024interplanetary} during geomagnetically active times. We are approaching the peak of the 25\textsuperscript{th} solar cycle even before reaching its maximum, a multi-CME generated intense geomagnetic storm; the strongest solar storm of two decades hit the magnetosphere on May 10 - 11, 2024. This event provides an opportunity to study complex plasma intricacies at Earth's magnetopause.

We report here the prolonged high-energy electron flux enhancement up to 650 keV and simultaneous observations of electron cyclotron harmonics, whistler, and Langmuir waves, notably without magnetic reconnection. These observations indicate the direct role of the electron-scale waves in heating the electrons at the magnetopause.
 
\section*{Results}

\subsection*{Dataset}

We use the data from the Magnetospheric Multiscale (MMS) spacecraft \cite{burch2016magnetospheric}. Specifically, high-resolution burst mode data is analyzed to study the electron wave dynamics. For three-dimensional electric field measurement, high-frequency AC coupled electric field (HMFE) data from electric field double probes (EDP) is analyzed \cite{ergun2016axial,lindqvist2016spin}. Magnetic field measurements are obtained using Fluxgate Magnetometer (FGM) \cite{russell2016magnetospheric} and Search Coil Magnetometer (SCM) \cite{le2016search}. FGM measures DC magnetic fields at a sampling frequency of 128 Hz, while EDP and SCM sample electric and magnetic field waveforms at 65.536 kHz and 16.384 kHz, respectively. For particle measurements, the data from Fast Plasma Investigation (FPI) \cite{pollock2016fast} and the Fly's Eye Energetic Electron Proton Spectrometer (FEEPS) \cite{blake2016fly} are used. The sampling frequency of burst mode FPI data is 33.33 Hz whereas, the FEEPS instrument can measure all-sky snapshots (nearly 360$^{\circ}$) of electron intensity from $\sim$25 to 650 keV every 0.33 s in burst mode. At this count, the local magnetic field measurements can be used to determine pitch angle distributions at differential energies. To provide an overview of localized magnetospheric parameters, fast and survey mode datasets are used. The burst dataset of the electric field is not available for MMS 4. Other than this all the datasets mentioned above are available for the other three spacecraft. In this paper, the observations of the MMS 3 satellite are shown. The geomagnetic index SYM-H and other interplanetary parameters are obtained from the OMNI dataset.

\subsection*{Prolonged High-Energy Flux Enhancement at the Magnetopause}

We report the first ever observation of prolonged high-energy electron flux enhancements up to 650 keV during the MMS's magnetopause crossing, notably without magnetic reconnection signatures. It is also possible that this flux enhancement could exceed 650 keV; however, this cannot be confirmed as the maximum energy coverage limit of the FEEPS instrument is capped at 650 keV. These observations are during the recovery phase of the severe G5 storm which occurred on May 10 - 11, 2024 as shown in the  in the Supplementary Figure 1.

\begin{figure}[h]
\includegraphics[width=0.8\linewidth, keepaspectratio=true]{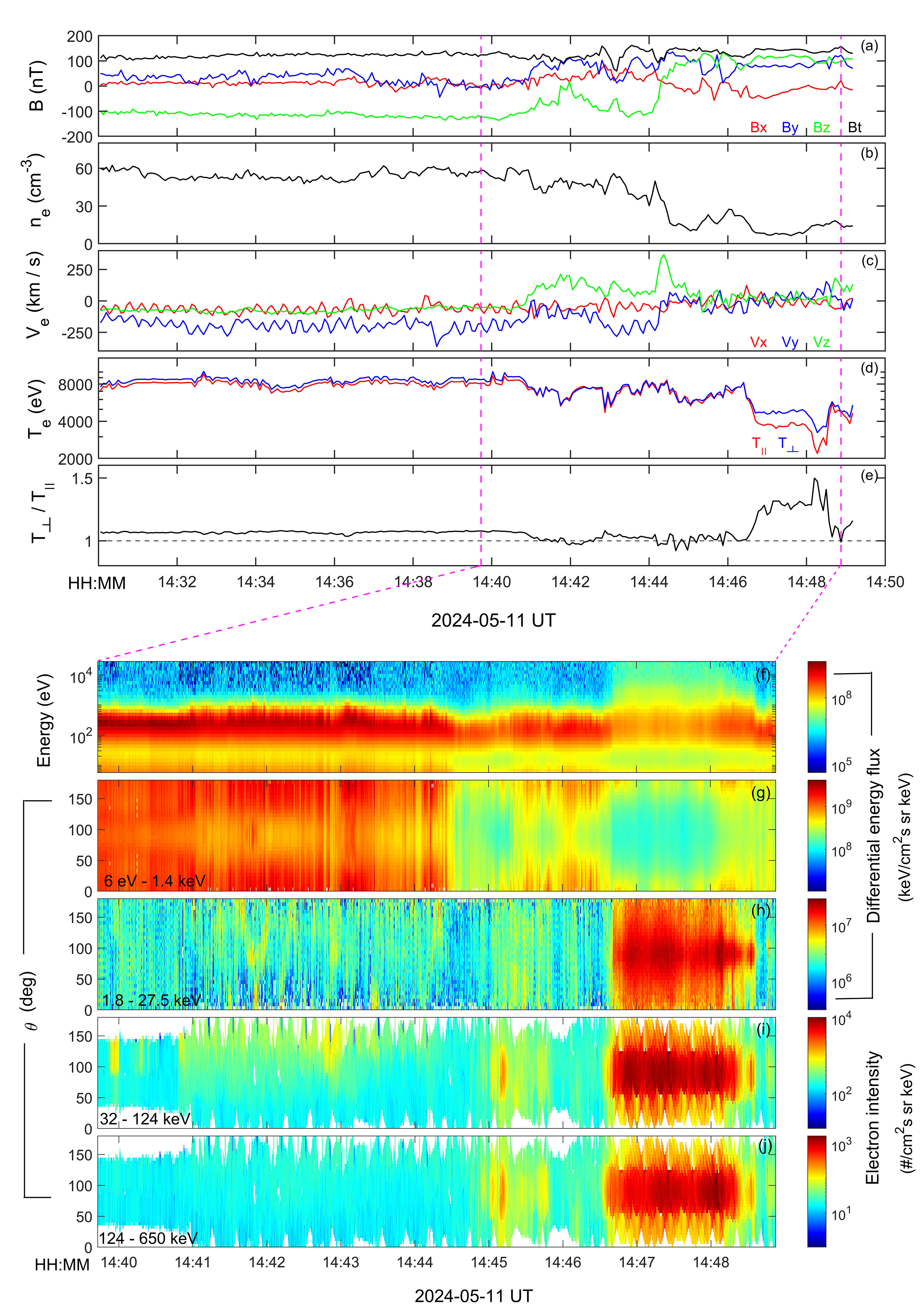}
\caption {MMS parameters: (a) Magnetic field in GSE coordinate system, (b) Electron density, (c) Electron velocity, (d) Electron temperature for energy range 1 keV to 27.5 keV, (e) Temperature anisotropy for 1 keV to 27.5 keV. Magenta dashed lines indicate the availability of the burst mode dataset, (f) Electron omnidirectional energy spectrogram, (g) Pitch angle distribution of 6 eV - 1.4 keV electrons, (h) Pitch angle distribution of 1.8 - 27.5 keV electrons, (i) Pitch angle distribution of 32 - 124 keV electrons, (j) Pitch angle distribution of 124 - 650 keV electrons.}
\label{fig:Fig1}
\end{figure}

On May 11, 2024, at 14:39 UT MMS was situated around [6.1465, -3.0864,-0.0602]$R_{E}$, where $R_{E}$ is the Earth's radius in the GSE coordinate system. The maximum separation between MMS spacecrafts is around 75 km. The trajectory of the MMS spacecraft orbit and their tetrahedral geometry is shown in Supplementary Figure 2. The red dashed lines in the Supplementary Figure 1 mark the timestamps when MMS was positioned near the subsolar point at the magnetopause during the recovery phase of a geomagnetic storm. An overview of the event, the electron measurements including electron density (n$_e$), velocity components (v$_x$, v$_y$, v$_z$), parallel and perpendicular temperature (T$_\parallel$, T$_\perp$) along with magnetic field components (B$_x$, B$_y$, B$_z$) is shown in Figure \ref{fig:Fig1}a-e. The magenta dashed lines in this figure highlight the periods when burst mode data is available. Focusing on this specific region, we observe fluctuations in the magnetospheric magnetic field (Figure \ref{fig:Fig1}a), density gradient (Figure \ref{fig:Fig1}b), parallel and perpendicular temperature (Figure \ref{fig:Fig1}d), temperature anisotropy in the perpendicular direction and its intensification (Figure \ref{fig:Fig1}e). These factors may act as a precursor for the generation of electron-scale waves locally.

Figure \ref{fig:Fig1}f shows the omnidirectional electron flux and the electron flux across different energy ranges across all pitch angles 0$^\circ$ to 180$^\circ$: low-energy 6 eV - 1.4 keV (Figure \ref{fig:Fig1}g), mid-energy 1.8 - 27.5 keV (Figure \ref{fig:Fig1}h), high-energy 32 - 124 keV (Figure \ref{fig:Fig1}i) and 124 - 650 keV (Figure \ref{fig:Fig1}j). Initially, there is a higher differential energy flux (DEF) in the low-energy range (Figure \ref{fig:Fig1}g), with moderate fluxes in the mid (Figure \ref{fig:Fig1}h) and high-energy ranges (Figure \ref{fig:Fig1}i, j). However, these moderate and high energy fluxes get significantly intensified, termed as the high energy electron flux enhancement (Figure \ref{fig:Fig1}h, i, j).
Omnidirectional DEF increases (Figure \ref{fig:Fig1}f) and is coincident with a decrease in the DEF of low-energy (Figure \ref{fig:Fig1}g) and an increase in the DEF of mid-energy (Figure \ref{fig:Fig1}h) and differential particle flux (DPF) of high-energy electrons (Figure \ref{fig:Fig1}i, j).

\subsection*{Electron-Scale Waves at High-Energy Flux Enhancement}

\begin{figure}[h]
\centering
\includegraphics[width=1\linewidth]{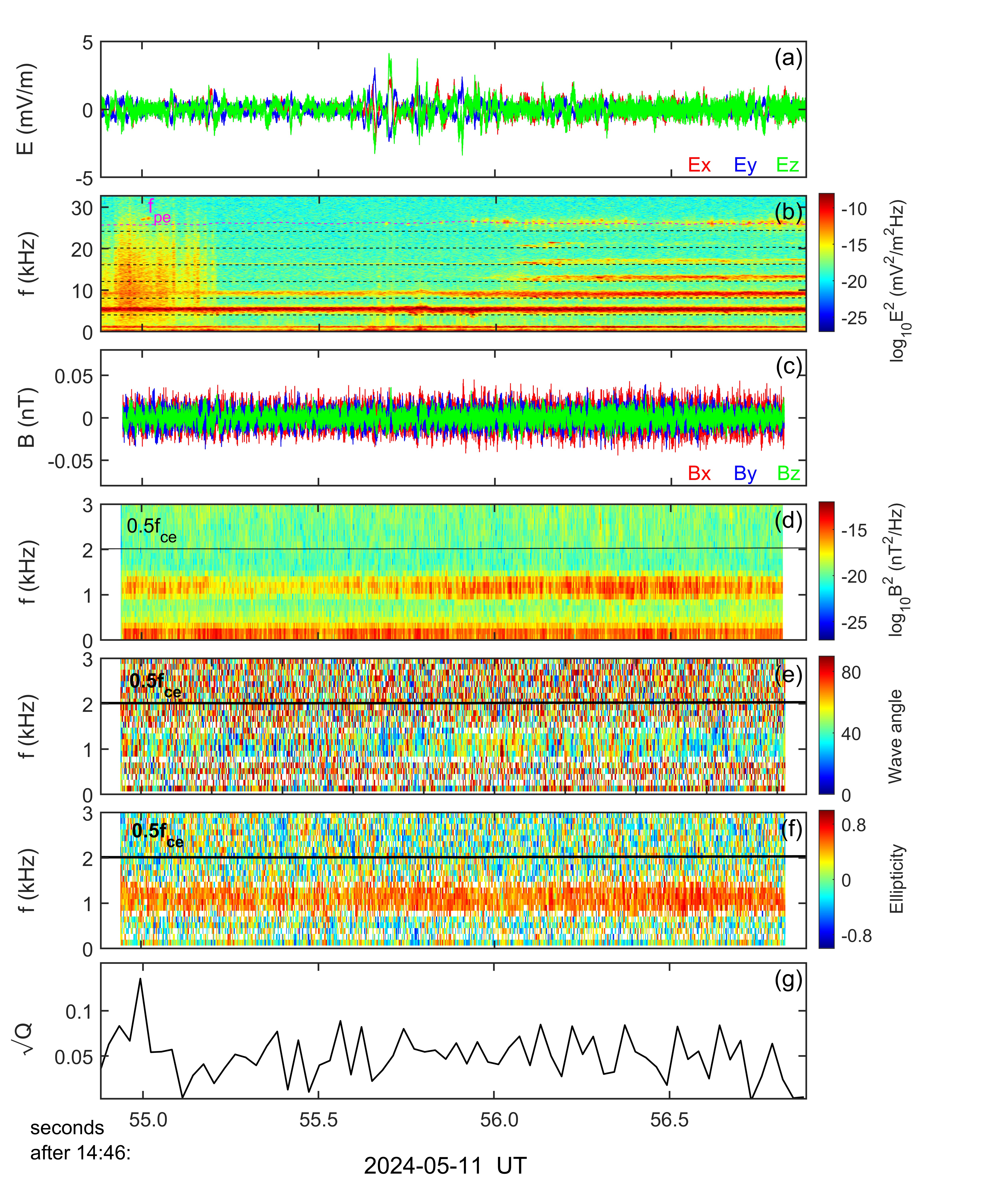}
\caption{Wave event 1: Simultaneous observation of whistler wave, electron cyclotron harmonics (ECH), and Langmuir wave: (a) Electric field in GSE coordinate system, (b) Electric field spectrogram (black and magenta dashed lines represent n$f_{ce}$, where n = 1 to 6 and $f_{pe}$ respectively), (c) Magnetic field in GSE coordinate system, (d) Magnetic field spectrogram, (e) Wave normal angle of whistler wave, (f) Ellipticity of whistler wave (solid black lines in panel d, e, f represent 0.5$f_{ce}$), (g) Agyrotropy parameter for 1 keV to 27.5 keV electrons.}
\label{fig:Fig2}
\end{figure}

\begin{figure}[h]
\includegraphics[width=1\linewidth]{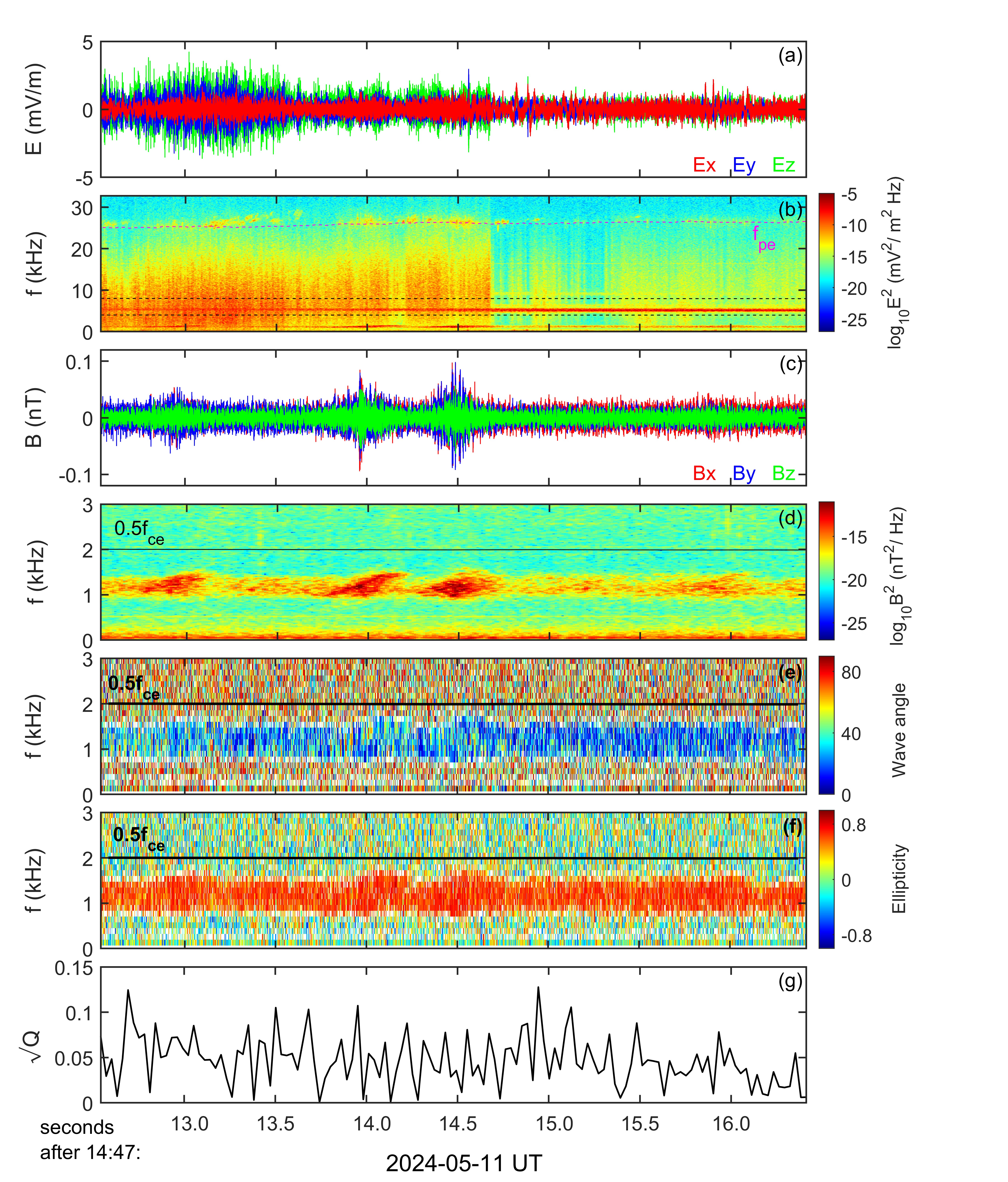}
\caption{Wave event 2: Simultaneous observation of whistler wave, electron cyclotron harmonics (ECH), and Langmuir wave: (a) Electric field in GSE coordinate system, (b) Electric field spectrogram (black and magenta dashed lines represent n$f_{ce}$, where n = 1, 2 and $f_{pe}$ respectively), (c) Magnetic field in GSE coordinate system, (d) Magnetic field spectrogram, (e) Wave normal angle of whistler wave, (f) Ellipticity of whistler wave (solid black lines in panel d, e, f represent 0.5$f_{ce}$), (g) Agyrotropy parameter for 1 keV to 27.5 keV electrons. }
\label{fig:Fig3}
\end{figure}

We investigate the wave emissions during this high energy flux enhancement to examine whether the energetic electrons were produced locally by wave-particle interaction processes. Figures \ref{fig:Fig2} and \ref{fig:Fig3} show the simultaneous electron wave events observed during high-energy flux enhancement. Figure \ref{fig:Fig2} illustrates the period from 14:46:54.882 UT to 14:46:56.882 UT. The frequency of the Hiss-like whistler band around 1 kHz, which is less than 0.5$f_{ce}$ is identified from the electric and magnetic field spectrograms (Figure \ref{fig:Fig2}b, d). The wave normal angle is between 10$^{\circ}$ to 40$^{\circ}$, indicating their quasi-parallel propagation (Figure \ref{fig:Fig2}e). The ellipticity is close to 1, implying their right-handed circular polarization (Figure \ref{fig:Fig2}f) that confirms the whistler wave. The electric field spectrogram shows a fundamental ECH band frequency around 4 kHz, with four harmonics at frequencies that are multiples of (n+1)$f_{ce}$, where $n$ ranges from 1 to 4, extending up to electron plasma frequency, $f_{pe}$ around 26 kHz (Figure \ref{fig:Fig2}b). The wave at $f_{pe}$ is identified as the Langmuir wave by using the $F_{E}$ index analysis proposed by Graham et al.\cite{graham2018large}, where $F_{E}$ is the ratio of perpendicular power to total power in the snapshot, which is given by $F_{E} $ = $\frac{\sum E^2_{\perp}}{\sum E^2_{\perp} + \sum E^2_{\parallel}}$
. For Langmuir waves, $F_{E}$ $<$ 0.5. For this case, $F_{E}$ = 0.092 confirms that the observed wave is a Langmuir wave. Figure \ref{fig:Fig2}g shows an agyrotropy parameter \cite{swisdak2016quantifying} which quantifies the agyrotropy of electrons, measuring how much the electron velocity distribution deviates from gyrotropy. The larger the $Q$ value, the stronger the electron agyrotropy. 

Another event of the simultaneous observation of whistler, ECH, and Langmuir waves from 14:47:12.537 UT to 14:47:16.416 UT is shown in Figure \ref{fig:Fig3}. The frequency of the whistler wave is approximately 1 kHz and is less than 0.5$f_{ce}$ (Figure \ref{fig:Fig3}b, d), unlike the Hiss-like band, displays rising tones, aligning with intensifying magnetic field fluctuations (Figure \ref{fig:Fig3}c). The wave normal angle is around 10$^{\circ}$ (Figure \ref{fig:Fig3}e), and the ellipticity is close to 1 (Figure \ref{fig:Fig3}f). The electric field spectrogram shows both the fundamental band and the first harmonic of the ECH wave with frequencies 5 kHz and 10 kHz, corresponding to (n+1)$f_{ce}$, respectively (Figure \ref{fig:Fig3}b). The wave at frequency $f_{pe}$, around 25 kHz is identified as a Langmuir wave as the $F_{E}$ index obtained here is 0.1\cite{graham2018large} (Figure \ref{fig:Fig3}b). Figure \ref{fig:Fig3}g shows the agyrotropy parameter during the wave event 2 (Figure \ref{fig:Fig2}g).

During this electron flux intensification, 16 bursts were observed in electric and magnetic field datasets. Among these, 8 bursts observed simultaneous wave activity of whistler waves, ECH waves, and Langmuir waves; 4 bursts showed simultaneous ECH waves and Langmuir waves; 1 burst contained both whistler and ECH waves, and 1 burst contained only ECH wave. 

\subsection*{Generation Mechanism of the Waves}
\begin{figure}[h]
\includegraphics[width=1\linewidth]{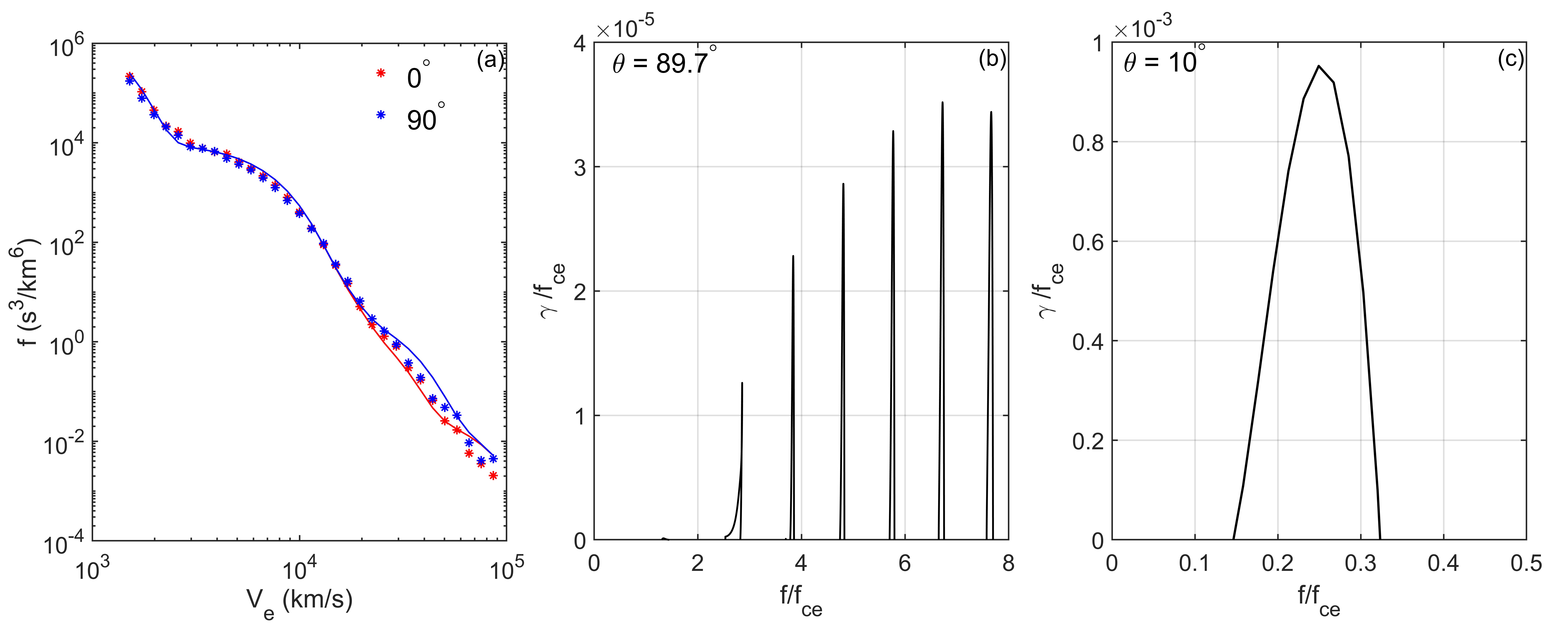}
\caption{The electron distributions and growth rates for wave event 1: (a) Electron phase space density vs electron velocity in parallel and perpendicular directions with red and blue representing 0$^{\circ}$ and 90$^{\circ}$ respectively, at 14:46:55.682 UT, (b) The growth rate of ECH and Langmuir wave at 89.7$^{\circ}$, (c) The growth rate of whistler wave at 10$^{\circ}$.}
\label{fig:Fig4}
\end{figure}

The observed electron distributions in the parallel and perpendicular directions are fitted using a subtracted bi-Maxwellian distribution with loss cone parameters (Figure \ref{fig:Fig4}a) (refer to the Electron Velocity Distribution Functions in the Methods section). These fitting parameters are given in Table\ref{table1}. They were then used as input to the dispersion solver KUPDAP (Kyoto University Plasma Dispersion Analysis Package) \cite{sugiyama2015electromagnetic} to analyze the growth of the instabilities (refer to the Linear Dispersion Analysis in the Methods section). For wave event 1 (Figure \ref{fig:Fig2}), the growth of the ECH wave and whistler wave is observed at 89.7$^{\circ}$ and 10$^{\circ}$, respectively (Figure \ref{fig:Fig4}b, c). The growth of the fundamental band of waves around the $f_{ce}$ and for harmonics around (n+1)$f_{ce}$, as well as the Langmuir waves is seen around 27 kHz (Figure \ref{fig:Fig4}b). This is in close agreement with the observed results. This analysis concludes that the observed background plasma conditions support the growth of instabilities that drive the observed electron-scale waves. This hints that the waves are locally generated by the electron loss cone distribution and their simultaneous occurrence is also verified using linear dispersion analysis. A similar analysis to check the growth of the instabilities is done for wave event 2 (Figure \ref{fig:Fig3}). This case also shows the growth of the whistler wave at 10$^{\circ}$ and ECH waves at 89.7$^{\circ}$, which is not shown here.

\section*{Discussion}

Although energetic electrons up to 125 keV have been observed at the magnetopause, such observations have not been observed frequently \cite{jaynes2016energetic,fu2019evidence,chepuri2022statistical}. The statistical study of 250 magnetopause crossing of MMS shows either a direct or indirect connection between energetic electron enhancement and reconnection \cite{chepuri2022statistical}. Whistler waves and energetic electrons are believed to be frequent outcomes of magnetic reconnection. However, more than half of the events in the statistical study observed energetic electrons without the occurrence of whistler waves \cite{chepuri2022statistical}.

We report here the energetic electron flux enhancement up to 650 keV, which has no link to magnetic reconnection. In addition to whistler waves, the energetic electron flux event reported here exhibits Langmuir waves and ECH waves. Such a simultaneous observation of whistler waves, Langmuir waves, and ECH waves along with the prolonged electron flux enhancement near the Earth's magnetopause crossing has never been reported. These observations demonstrate the direct association of electron flux enhancement with the observed electron-scale waves. 

The energy exchange parameter \( \vec{J} \cdot \vec{E} \) is computed for wave event 1, i.e. (Figure \ref{fig:Fig2}), and positive values for the whistler wave (0.0120 \,$ \text{nW/m}^3$), the fundamental band of the ECH wave (0.1781 \,$ \text{nW/m}^3$), and the Langmuir wave (0.0004 \,$ \text{nW/m}^3$) are obtained as shown in the Supplementary Figure 3. This indicates a wave-particle interaction, suggesting that the electron-scale waves are transferring energy to the electrons \cite{liu2024ultrafast} (refer to the Energy Transfer Parameter in Methods section). The electric field power of these waves is \(1919 \, (\text{mV/m})^2 \), \( 19343 \, (\text{mV/m})^2 \), and \(26.43 \, (\text{mV/m})^2 \), respectively, highlighting the ECH wave may play a dominant role in energy exchange.

Our observations address two questions raised by Graham et al. \cite{graham2018large}: what instabilities generate waves around $f_{pe}$ and do these waves contribute to plasma heating at magnetopause? The electron loss cone distribution functions, that drive electron-scale instabilities, are found simultaneously with the electron flux enhancement. Temperature anisotropy and the loss cone electron distributions observed at the magnetopause are likely to be the source of whistler waves, electron cyclotron harmonics waves, and Langmuir waves. The linear kinetic dispersion analysis shows that loss cone particle distribution provides a free energy source to excite these electron-scale instabilities that drive the respective wave. The observed high energetic electron flux event demonstrates the results of a cascading of all the possible effects via whistler waves, electron cyclotron harmonic waves and Langmuir waves. It is difficult to reveal the quantitative effects of each observed wave separately from the data. Further numerical simulations using particle-in-cell models are needed to quantify the systematic combined effects of whistler, electron cyclotron, and Langmuir waves from linear to nonlinear stages. Such a simulation would bring out the crucial role of cascading wave-particle interaction in understanding the electron energization and prolonged flux enhancement at the magnetopause, which has large-scale implications in near-Earth space.

\section*{Methods}

\subsection*{Electron Velocity Distribution Functions}
We fit the observed electron distributions (Figure \ref{fig:Fig4}a) by the subtracted bi-Maxwellian equation:
\begin{equation}
f(v_\parallel, v_\perp) = \frac{n}{\pi^{3/2} v_{\mathrm{th}\parallel} v_{\mathrm{th}\perp}^2} \exp\left( - \frac{(v_\parallel - v_d)^2}{v_{\mathrm{th}\parallel}^2} \right) 
\left\{ \Delta \exp\left( -\frac{v_\perp^2}{v_{\mathrm{th}\perp}^2} \right) + \frac{1 - \Delta}{1 - \beta} 
\left[ \exp\left( -\frac{v_\perp^2}{v_{\mathrm{th}\perp}^2} \right) - \exp\left( -\frac{v_\perp^2}{\beta v_{\mathrm{th}\perp}^2} \right) \right] 
\right\}
\end{equation}
where, $v_{\text{th}\parallel}$ and $v_{\text{th}\perp}$ are the parallel and perpendicular thermal velocities respectively calculated using the parallel and perpendicular temperature as $v_{\text{th}} = \sqrt{\frac{2kT}{m_e}}$. $k$ is the Boltzmann constant,  $m_e$ is the electron mass, $n$ is the density, $v_d$ is the drift velocity, $v_\parallel$ and $v_\perp$ are the parallel and perpendicular velocity components,
$\Delta$ and $\beta$ are the loss cone parameters representing the loss cone depth and width, respectively.

The fitting parameters are given in Table\ref{table1}. Only electron components from Table\ref{table1} are used to fit the observed velocity distribution (Figure \ref{fig:Fig4}a). The ion component is included to ensure plasma quasineutrality in KUPDAP. The drift velocity $v_d$ is zero for all components.

\subsection*{Linear Dispersion Analysis}
The fitting parameters in Table\ref{table1} are used in the Kyoto University Plasma Dispersion Analysis Package (KUPDAP) to investigate the growth of the instabilities. This Package is developed by the Space Group at the Kyoto University, Japan, is a tool for solving dispersion relations under the assumption of an infinite, uniform, collisionless plasma composed of multiple components, embedded in a consistent magnetic field\cite{sugiyama2015electromagnetic}.
We use a loss cone distribution in the form of a subtracted bi-Maxwellian with loss cone parameters to model observed electron waves and assess whether the background plasma conditions support the growth of instabilities. In KUPDAP, the subtracted bi-Maxwellian distribution with loss cone parameters is applied to the anisotropic electron population, while all other electrons and ions are represented by a standard bi-Maxwellian distribution, which simplifies to a Maxwellian without any loss cone parameters.
\subsection*{Energy Transfer Parameter}
The energy exchange within the waves and particles is quantified by \( \vec{J} \cdot \vec{E} \), where \( \vec{E} \) represents the wave electric field and \( \vec{J}(n_e, v_e) \) denotes turbulent electron currents generated. This parameter characterizes wave-particle interactions. The parameter \( \vec{J} \cdot \vec{E} \) remains predominantly positive  indicating the continuous energy transfer from the waves to the electrons and vice versa \cite{liu2024ultrafast}.

\subsection*{Data availability}

MMS level-2 data analyzed in the present study are publicly available via the MMS Science Data Center (https://lasp.colorado.edu/
mms/sdc/public/).

\subsection*{Code availability}

The Kyoto University Plasma Dispersion Analysis Package (KUPDAP) that was used to calculate the dispersion relation of
the electron-scale waves is available at http://space.rish.kyoto-u. ac.jp/software.

\bibliography{REFERENCES}

\begin{thebibliography}{10}
\urlstyle{rm}
\expandafter\ifx\csname url\endcsname\relax
  \def\url#1{\texttt{#1}}\fi
\expandafter\ifx\csname urlprefix\endcsname\relax\def\urlprefix{URL }\fi
\expandafter\ifx\csname doiprefix\endcsname\relax\def\doiprefix{DOI: }\fi
\providecommand{\bibinfo}[2]{#2}
\providecommand{\eprint}[2][]{\url{#2}}

\bibitem{frey2003continuous}
\bibinfo{author}{Frey, H.}, \bibinfo{author}{Phan, T.},
  \bibinfo{author}{Fuselier, S.} \& \bibinfo{author}{Mende, S.}
\newblock \bibinfo{journal}{\bibinfo{title}{Continuous magnetic reconnection at
  earth's magnetopause}}.
\newblock {\emph{\JournalTitle{Nature}}} \textbf{\bibinfo{volume}{426}},
  \bibinfo{pages}{533--537},
  \doiprefix\url{https://doi.org/10.1038/nature02084} (\bibinfo{year}{2003}).

\bibitem{hesse2020magnetic}
\bibinfo{author}{Hesse, M.} \& \bibinfo{author}{Cassak, P.}
\newblock \bibinfo{journal}{\bibinfo{title}{Magnetic reconnection in the space
  sciences: Past, present, and future}}.
\newblock {\emph{\JournalTitle{Journal of Geophysical Research: Space
  Physics}}} \textbf{\bibinfo{volume}{125}}, \bibinfo{pages}{e2018JA025935},
  \doiprefix\url{https://doi.org/10.1029/2018JA025935} (\bibinfo{year}{2020}).

\bibitem{hwang2015magnetopause}
\bibinfo{author}{Hwang, K.-J.}
\newblock \bibinfo{journal}{\bibinfo{title}{Magnetopause waves controlling the
  dynamics of earth’s magnetosphere}}.
\newblock {\emph{\JournalTitle{Journal of Astronomy and Space Sciences}}}
  \textbf{\bibinfo{volume}{32}}, \bibinfo{pages}{1--11},
  \doiprefix\url{https://doi.org/10.5140/JASS.2015.32.1.1}
  (\bibinfo{year}{2015}).

\bibitem{tang2013themis}
\bibinfo{author}{Tang, X.} \emph{et~al.}
\newblock \bibinfo{journal}{\bibinfo{title}{Themis observations of the
  magnetopause electron diffusion region: Large amplitude waves and heated
  electrons}}.
\newblock {\emph{\JournalTitle{Geophysical Research Letters}}}
  \textbf{\bibinfo{volume}{40}}, \bibinfo{pages}{2884--2890},
  \doiprefix\url{https://doi.org/10.1002/grl.50565} (\bibinfo{year}{2013}).

\bibitem{oieroset2002evidence}
\bibinfo{author}{{\O}ieroset, M.}, \bibinfo{author}{Lin, R.},
  \bibinfo{author}{Phan, T.}, \bibinfo{author}{Larson, D.} \&
  \bibinfo{author}{Bale, S.}
\newblock \bibinfo{journal}{\bibinfo{title}{Evidence for electron acceleration
  up to~ 300 ke v in the magnetic reconnection<? format?> diffusion region of
  earth’s magnetotail}}.
\newblock {\emph{\JournalTitle{Physical Review Letters}}}
  \textbf{\bibinfo{volume}{89}}, \bibinfo{pages}{195001},
  \doiprefix\url{https://doi.org/10.1103/PhysRevLett.89.195001}
  (\bibinfo{year}{2002}).

\bibitem{wang2010observations}
\bibinfo{author}{Wang, R.}, \bibinfo{author}{Lu, Q.}, \bibinfo{author}{Li, X.},
  \bibinfo{author}{Huang, C.} \& \bibinfo{author}{Wang, S.}
\newblock \bibinfo{journal}{\bibinfo{title}{Observations of energetic electrons
  up to 200 kev associated with a secondary island near the center of an ion
  diffusion region: A cluster case study}}.
\newblock {\emph{\JournalTitle{Journal of Geophysical Research: Space
  Physics}}} \textbf{\bibinfo{volume}{115}},
  \doiprefix\url{https://doi.org/10.1029/2010JA015473} (\bibinfo{year}{2010}).

\bibitem{zhang2024relativistic}
\bibinfo{author}{Zhang, X.-J.} \emph{et~al.}
\newblock \bibinfo{journal}{\bibinfo{title}{Relativistic and ultra-relativistic
  electron bursts in earth's magnetotail observed by low-altitude satellites}}.
\newblock {\emph{\JournalTitle{arXiv preprint arXiv:2408.17299}}}
  (\bibinfo{year}{2024}).

\bibitem{khotyaintsev2019collisionless}
\bibinfo{author}{Khotyaintsev, Y.~V.}, \bibinfo{author}{Graham, D.~B.},
  \bibinfo{author}{Norgren, C.} \& \bibinfo{author}{Vaivads, A.}
\newblock \bibinfo{journal}{\bibinfo{title}{Collisionless magnetic reconnection
  and waves: Progress review}}.
\newblock {\emph{\JournalTitle{Frontiers in Astronomy and Space Sciences}}}
  \textbf{\bibinfo{volume}{6}}, \bibinfo{pages}{70},
  \doiprefix\url{https://doi.org/10.3389/fspas.2019.00070}
  (\bibinfo{year}{2019}).

\bibitem{wei2007cluster}
\bibinfo{author}{Wei, X.} \emph{et~al.}
\newblock \bibinfo{journal}{\bibinfo{title}{Cluster observations of waves in
  the whistler frequency range associated with magnetic reconnection in the
  earth's magnetotail}}.
\newblock {\emph{\JournalTitle{Journal of Geophysical Research: Space
  Physics}}} \textbf{\bibinfo{volume}{112}},
  \doiprefix\url{https://doi.org/10.1029/2006JA011771} (\bibinfo{year}{2007}).

\bibitem{farrell2002dominance}
\bibinfo{author}{Farrell, W.}, \bibinfo{author}{Desch, M.},
  \bibinfo{author}{Kaiser, M.} \& \bibinfo{author}{Goetz, K.}
\newblock \bibinfo{journal}{\bibinfo{title}{The dominance of electron plasma
  waves near a reconnection x-line region}}.
\newblock {\emph{\JournalTitle{Geophysical research letters}}}
  \textbf{\bibinfo{volume}{29}}, \bibinfo{pages}{8--1},
  \doiprefix\url{https://doi.org/10.1029/2002GL014662} (\bibinfo{year}{2002}).

\bibitem{farrell2003role}
\bibinfo{author}{Farrell, W.}, \bibinfo{author}{Desch, M.},
  \bibinfo{author}{Ogilvie, K.}, \bibinfo{author}{Kaiser, M.} \&
  \bibinfo{author}{Goetz, K.}
\newblock \bibinfo{journal}{\bibinfo{title}{The role of upper hybrid waves in
  magnetic reconnection}}.
\newblock {\emph{\JournalTitle{Geophysical research letters}}}
  \textbf{\bibinfo{volume}{30}},
  \doiprefix\url{https://doi.org/10.1029/2003GL017549} (\bibinfo{year}{2003}).

\bibitem{jaynes2016energetic}
\bibinfo{author}{Jaynes, A.} \emph{et~al.}
\newblock \bibinfo{journal}{\bibinfo{title}{Energetic electron acceleration
  observed by mms in the vicinity of an x-line crossing}}.
\newblock {\emph{\JournalTitle{Geophysical Research Letters}}}
  \textbf{\bibinfo{volume}{43}}, \bibinfo{pages}{7356--7363},
  \doiprefix\url{https://doi.org/10.1002/2016GL069206} (\bibinfo{year}{2016}).

\bibitem{fu2019evidence}
\bibinfo{author}{Fu, H.} \emph{et~al.}
\newblock \bibinfo{journal}{\bibinfo{title}{Evidence of electron acceleration
  at a reconnecting magnetopause}}.
\newblock {\emph{\JournalTitle{Geophysical Research Letters}}}
  \textbf{\bibinfo{volume}{46}}, \bibinfo{pages}{5645--5652},
  \doiprefix\url{https://doi.org/10.1029/2019GL083032} (\bibinfo{year}{2019}).

\bibitem{chepuri2022statistical}
\bibinfo{author}{Chepuri, S.~N.} \emph{et~al.}
\newblock \bibinfo{journal}{\bibinfo{title}{A statistical study of magnetopause
  boundary layer energetic electron enhancements using mms}}.
\newblock {\emph{\JournalTitle{Frontiers in astronomy and space sciences}}}
  \textbf{\bibinfo{volume}{9}}, \bibinfo{pages}{926660},
  \doiprefix\url{https://doi.org/10.3389/fspas.2022.926660}
  (\bibinfo{year}{2022}).

\bibitem{contel2016whistler}
\bibinfo{author}{Contel, O.~L.} \emph{et~al.}
\newblock \bibinfo{journal}{\bibinfo{title}{Whistler mode waves and hall fields
  detected by mms during a dayside magnetopause crossing}}.
\newblock {\emph{\JournalTitle{Geophysical Research Letters}}}
  \textbf{\bibinfo{volume}{43}}, \bibinfo{pages}{5943--5952},
  \doiprefix\url{https://doi.org/10.1002/2016GL068968} (\bibinfo{year}{2016}).

\bibitem{wilder2016observations}
\bibinfo{author}{Wilder, F.} \emph{et~al.}
\newblock \bibinfo{journal}{\bibinfo{title}{Observations of whistler mode waves
  with nonlinear parallel electric fields near the dayside magnetic
  reconnection separatrix by the magnetospheric multiscale mission}}.
\newblock {\emph{\JournalTitle{Geophysical Research Letters}}}
  \textbf{\bibinfo{volume}{43}}, \bibinfo{pages}{5909--5917},
  \doiprefix\url{https://doi.org/10.1002/2016GL069473} (\bibinfo{year}{2016}).

\bibitem{zhou2018magnetospheric}
\bibinfo{author}{Zhou, M.} \emph{et~al.}
\newblock \bibinfo{journal}{\bibinfo{title}{Magnetospheric multiscale
  observations of an ion diffusion region with large guide field at the
  magnetopause: Current system, electron heating, and plasma waves}}.
\newblock {\emph{\JournalTitle{Journal of Geophysical Research: Space
  Physics}}} \textbf{\bibinfo{volume}{123}}, \bibinfo{pages}{1834--1852},
  \doiprefix\url{https://doi.org/10.1002/2017JA024517} (\bibinfo{year}{2018}).

\bibitem{zhou2016observation}
\bibinfo{author}{Zhou, M.} \emph{et~al.}
\newblock \bibinfo{journal}{\bibinfo{title}{Observation of high-frequency
  electrostatic waves in the vicinity of the reconnection ion diffusion region
  by the spacecraft of the magnetospheric multiscale (mms) mission}}.
\newblock {\emph{\JournalTitle{Geophysical Research Letters}}}
  \textbf{\bibinfo{volume}{43}}, \bibinfo{pages}{4808--4815},
  \doiprefix\url{https://doi.org/10.1002/2016GL069010} (\bibinfo{year}{2016}).

\bibitem{li2018local}
\bibinfo{author}{Li, J.} \emph{et~al.}
\newblock \bibinfo{journal}{\bibinfo{title}{Local excitation of whistler mode
  waves and associated langmuir waves at dayside reconnection regions}}.
\newblock {\emph{\JournalTitle{Geophysical Research Letters}}}
  \textbf{\bibinfo{volume}{45}}, \bibinfo{pages}{8793--8802},
  \doiprefix\url{https://doi.org/10.1029/2018GL078287} (\bibinfo{year}{2018}).

\bibitem{yu2021simultaneous}
\bibinfo{author}{Yu, X.} \emph{et~al.}
\newblock \bibinfo{journal}{\bibinfo{title}{Simultaneous observation of
  whistler waves and electron cyclotron harmonic waves in the separatrix region
  of magnetopause reconnection}}.
\newblock {\emph{\JournalTitle{Journal of Geophysical Research: Space
  Physics}}} \textbf{\bibinfo{volume}{126}}, \bibinfo{pages}{e2021JA029609},
  \doiprefix\url{https://doi.org/10.1029/2021JA029609} (\bibinfo{year}{2021}).

\bibitem{peng2020variation}
\bibinfo{author}{Peng, Q.} \emph{et~al.}
\newblock \bibinfo{journal}{\bibinfo{title}{Variation of dayside chorus waves
  associated with solar wind dynamic pressure based on mms observations}}.
\newblock {\emph{\JournalTitle{Advances in Space Research}}}
  \textbf{\bibinfo{volume}{65}}, \bibinfo{pages}{2551--2558},
  \doiprefix\url{https://doi.org/10.1016/j.asr.2020.03.006}
  (\bibinfo{year}{2020}).

\bibitem{meredith2009survey}
\bibinfo{author}{Meredith, N.~P.}, \bibinfo{author}{Horne, R.~B.},
  \bibinfo{author}{Thorne, R.~M.} \& \bibinfo{author}{Anderson, R.~R.}
\newblock \bibinfo{journal}{\bibinfo{title}{Survey of upper band chorus and ech
  waves: Implications for the diffuse aurora}}.
\newblock {\emph{\JournalTitle{Journal of Geophysical Research: Space
  Physics}}} \textbf{\bibinfo{volume}{114}},
  \doiprefix\url{https://doi.org/10.1029/2009JA014230} (\bibinfo{year}{2009}).

\bibitem{ni2017statistical}
\bibinfo{author}{Ni, B.}, \bibinfo{author}{Gu, X.}, \bibinfo{author}{Fu, S.},
  \bibinfo{author}{Xiang, Z.} \& \bibinfo{author}{Lou, Y.}
\newblock \bibinfo{journal}{\bibinfo{title}{A statistical survey of
  electrostatic electron cyclotron harmonic waves based on themis fff wave
  data}}.
\newblock {\emph{\JournalTitle{Journal of Geophysical Research: Space
  Physics}}} \textbf{\bibinfo{volume}{122}}, \bibinfo{pages}{3342--3353},
  \doiprefix\url{https://doi.org/10.1002/2016JA023433} (\bibinfo{year}{2017}).

\bibitem{liu2020global}
\bibinfo{author}{Liu, X.}, \bibinfo{author}{Chen, L.}, \bibinfo{author}{Engel,
  M.~A.} \& \bibinfo{author}{Jordanova, V.~K.}
\newblock \bibinfo{journal}{\bibinfo{title}{Global simulation of electron
  cyclotron harmonic wave instability in a storm-time magnetosphere}}.
\newblock {\emph{\JournalTitle{Geophysical Research Letters}}}
  \textbf{\bibinfo{volume}{47}}, \bibinfo{pages}{e2019GL086368},
  \doiprefix\url{https://doi.org/10.1029/2019GL086368} (\bibinfo{year}{2020}).

\bibitem{xie2024interplanetary}
\bibinfo{author}{Xie, Y.} \emph{et~al.}
\newblock \bibinfo{journal}{\bibinfo{title}{Interplanetary shock induced
  intensification of electron cyclotron harmonic waves in the earth’s inner
  magnetosphere}}.
\newblock {\emph{\JournalTitle{Frontiers in Physics}}}
  \textbf{\bibinfo{volume}{12}}, \bibinfo{pages}{1334531},
  \doiprefix\url{https://doi.org/10.3389/fphy.2024.1334531}
  (\bibinfo{year}{2024}).

\bibitem{burch2016magnetospheric}
\bibinfo{author}{Burch, J.}, \bibinfo{author}{Moore, T.},
  \bibinfo{author}{Torbert, R.} \& \bibinfo{author}{Giles, B.-h.}
\newblock \bibinfo{journal}{\bibinfo{title}{Magnetospheric multiscale overview
  and science objectives}}.
\newblock {\emph{\JournalTitle{Space Science Reviews}}}
  \textbf{\bibinfo{volume}{199}}, \bibinfo{pages}{5--21},
  \doiprefix\url{https://doi.org/10.1007/s11214-015-0164-9}
  (\bibinfo{year}{2016}).

\bibitem{ergun2016axial}
\bibinfo{author}{Ergun, R.} \emph{et~al.}
\newblock \bibinfo{journal}{\bibinfo{title}{The axial double probe and fields
  signal processing for the mms mission}}.
\newblock {\emph{\JournalTitle{Space Science Reviews}}}
  \textbf{\bibinfo{volume}{199}}, \bibinfo{pages}{167--188},
  \doiprefix\url{https://doi.org/10.1007/s11214-014-0115-x}
  (\bibinfo{year}{2016}).

\bibitem{lindqvist2016spin}
\bibinfo{author}{Lindqvist, P.-A.} \emph{et~al.}
\newblock \bibinfo{journal}{\bibinfo{title}{The spin-plane double probe
  electric field instrument for mms}}.
\newblock {\emph{\JournalTitle{Space Science Reviews}}}
  \textbf{\bibinfo{volume}{199}}, \bibinfo{pages}{137--165},
  \doiprefix\url{https://doi.org/10.1007/s11214-014-0116-9}
  (\bibinfo{year}{2016}).

\bibitem{russell2016magnetospheric}
\bibinfo{author}{Russell, C.} \emph{et~al.}
\newblock \bibinfo{journal}{\bibinfo{title}{The magnetospheric multiscale
  magnetometers}}.
\newblock {\emph{\JournalTitle{Space Science Reviews}}}
  \textbf{\bibinfo{volume}{199}}, \bibinfo{pages}{189--256},
  \doiprefix\url{https://doi.org/10.1007/s11214-014-0057-3}
  (\bibinfo{year}{2016}).

\bibitem{le2016search}
\bibinfo{author}{Le~Contel, O.} \emph{et~al.}
\newblock \bibinfo{journal}{\bibinfo{title}{The search-coil magnetometer for
  mms}}.
\newblock {\emph{\JournalTitle{Space Science Reviews}}}
  \textbf{\bibinfo{volume}{199}}, \bibinfo{pages}{257--282},
  \doiprefix\url{https://doi.org/10.1007/s11214-014-0096-9}
  (\bibinfo{year}{2016}).

\bibitem{pollock2016fast}
\bibinfo{author}{Pollock, C.} \emph{et~al.}
\newblock \bibinfo{journal}{\bibinfo{title}{Fast plasma investigation for
  magnetospheric multiscale}}.
\newblock {\emph{\JournalTitle{Space Science Reviews}}}
  \textbf{\bibinfo{volume}{199}}, \bibinfo{pages}{331--406},
  \doiprefix\url{https://doi.org/10.1007/s11214-016-0245-4}
  (\bibinfo{year}{2016}).

\bibitem{blake2016fly}
\bibinfo{author}{Blake, J.} \emph{et~al.}
\newblock \bibinfo{journal}{\bibinfo{title}{The fly’s eye energetic particle
  spectrometer (feeps) sensors for the magnetospheric multiscale (mms)
  mission}}.
\newblock {\emph{\JournalTitle{Space Science Reviews}}}
  \textbf{\bibinfo{volume}{199}}, \bibinfo{pages}{309--329},
  \doiprefix\url{https://doi.org/10.1007/s11214-015-0163-x}
  (\bibinfo{year}{2016}).

\bibitem{graham2018large}
\bibinfo{author}{Graham, D.~B.} \emph{et~al.}
\newblock \bibinfo{journal}{\bibinfo{title}{Large-amplitude high-frequency
  waves at earth's magnetopause}}.
\newblock {\emph{\JournalTitle{Journal of Geophysical Research: Space
  Physics}}} \textbf{\bibinfo{volume}{123}}, \bibinfo{pages}{2630--2657},
  \doiprefix\url{https://doi.org/10.1002/2017JA025034} (\bibinfo{year}{2018}).

\bibitem{swisdak2016quantifying}
\bibinfo{author}{Swisdak, M.}
\newblock \bibinfo{journal}{\bibinfo{title}{Quantifying gyrotropy in magnetic
  reconnection}}.
\newblock {\emph{\JournalTitle{Geophysical Research Letters}}}
  \textbf{\bibinfo{volume}{43}}, \bibinfo{pages}{43--49},
  \doiprefix\url{https://doi.org/10.1002/2015GL066980} (\bibinfo{year}{2016}).

\bibitem{sugiyama2015electromagnetic}
\bibinfo{author}{Sugiyama, H.} \emph{et~al.}
\newblock \bibinfo{journal}{\bibinfo{title}{Electromagnetic ion cyclotron waves
  in the earth's magnetosphere with a kappa-maxwellian particle distribution}}.
\newblock {\emph{\JournalTitle{Journal of Geophysical Research: Space
  Physics}}} \textbf{\bibinfo{volume}{120}}, \bibinfo{pages}{8426--8439},
  \doiprefix\url{https://doi.org/10.1002/2015JA021346} (\bibinfo{year}{2015}).

\bibitem{liu2024ultrafast}
\bibinfo{author}{Liu, C.} \emph{et~al.}
\newblock \bibinfo{journal}{\bibinfo{title}{Ultrafast measurement of
  field-particle energy transfer during chorus emissions in space}}.
\newblock {\emph{\JournalTitle{arXiv preprint arXiv:2408.13156}}}
  (\bibinfo{year}{2024}).

\end{thebibliography}



\section*{Acknowledgments}

We thank the MMS Science Data Center (https://lasp.colorado.edu/mms/sdc/public/) for providing the data for this study. The authors thank the entire MMS team and instrument PIs for data access and support. S.L., A.K., and B.K. acknowledge support from IIG for research funding under the MI-PEARL project.

\section*{Author contributions statement}

A.K. designed the study and oversaw the project. S.L. carried out the data analysis under the supervision of A.K. and prepared the initial draft of the manuscript. A.K. and B.K. contributed to the data interpretation and edited the manuscript.

\section*{Competing interests}

The authors declare no competing interests.

\begin{table}[h]
\centering
\begin{tabular}{|c|c|c|c|c|c|}
\hline
Components & n $(\text{cm}^{-3})$ & $T_{\parallel}$ (eV) & $T_{\perp}$ (eV) & $\Delta$ & $\beta$ \\ 
\hline
1.  Electron & 15    & 5     & 5     & 0   & 0   \\ 
2.  Electron & 10.5  & 184   & 184   & 0   & 0   \\ 
3.  Electron & 1     & 500   & 500   & 0   & 0   \\ 
4.  Electron & 0.3   & 2000  & 3200  & 1   & 0.2   \\ 
5.  Electron & 0.05  & 1.95 $\times 10^4 $
 & 1.95 $\times 10^4 $
  & 0   & 0 \\ 
6.  Proton   & 26.85 & 2000  & 2000  & 0   & 0   \\ 
\hline
\end{tabular}
\caption{Plasma parameters used in fitting the observed electron distribution shown in Figure \ref{fig:Fig4}a}
\label{table1}
\end{table}

\clearpage

\section*{Supplementary information}
\begin{figure}[h]
\includegraphics[width=1\linewidth]{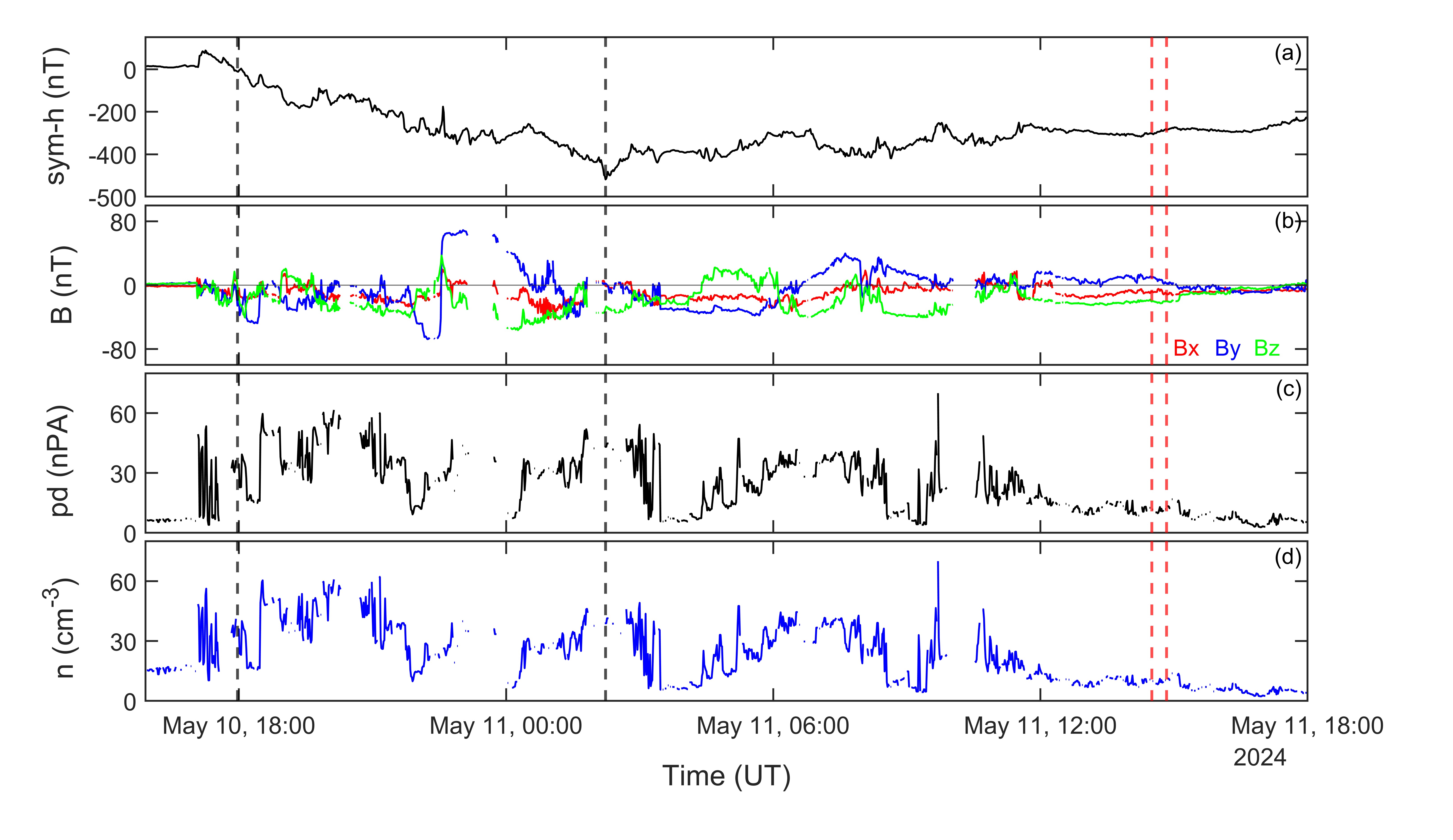}

\caption*{\textbf{Supplementary Figure 1}: The strongest storm of two decades is recorded in OMNI data: (a) SYM-H, (b) Interplanetary magnetic field in GSE coordinate system, (c) Solar wind dynamic pressure, (d) Proton density. Black dashed lines indicate the timestamps corresponding to the storm's main phase and red dashed lines indicate the Magnetospheric Multiscale (MMS) satellites were at the magnetopause.}
\label{fig:SupplementaryFig1}
\end{figure}
 
\begin{figure}[h]
\includegraphics[width=1\linewidth]{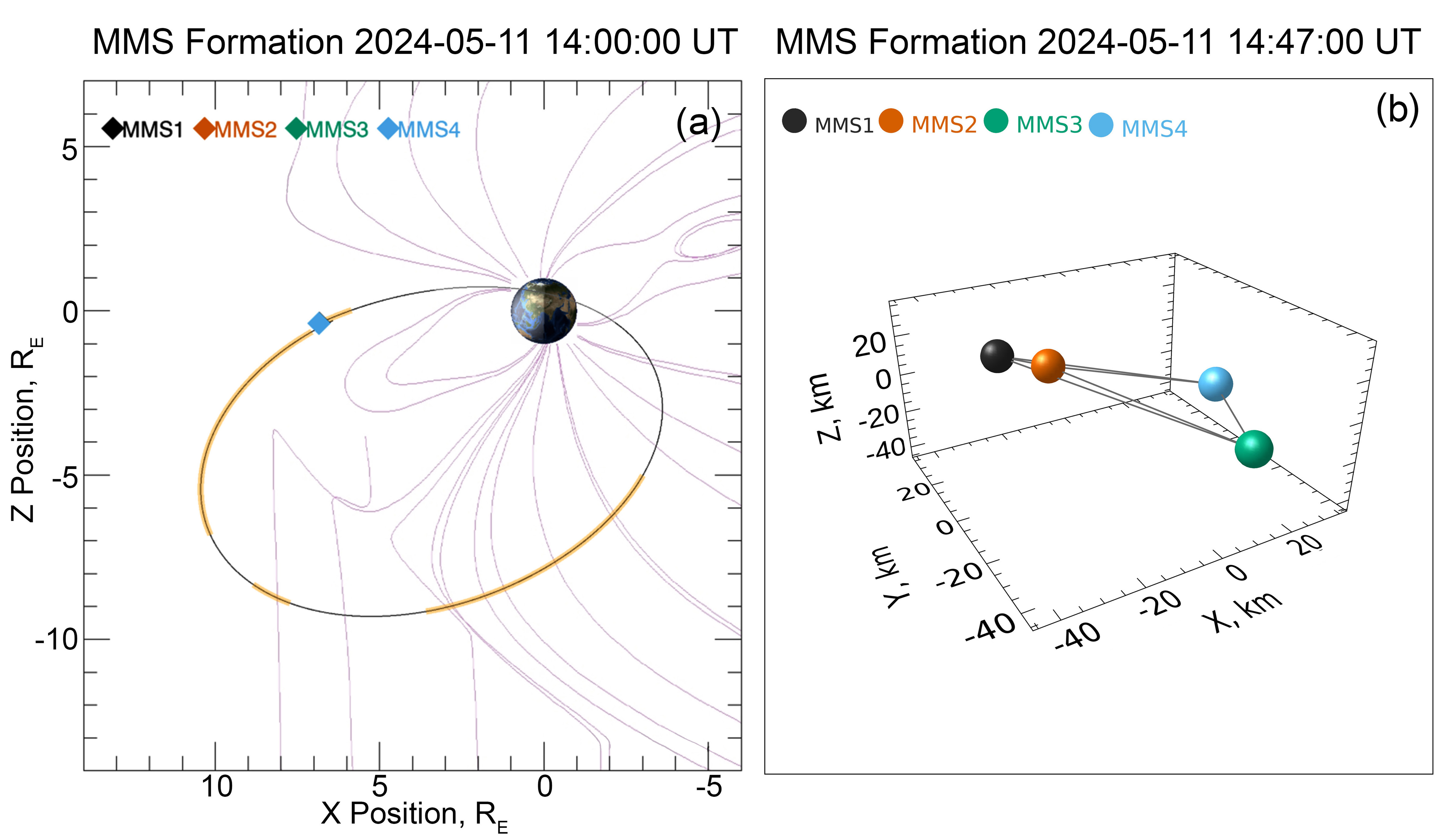}

\caption*{\textbf{Supplementary Figure 2}: {MMS parameters: (a) MMS around magnetopause, orbiting towards Earth, (b) Tetrahedral geometry of all MMS satellites.}}
\label{fig:SupplementaryFig2}
\end{figure}

\begin{figure}[h]
\includegraphics[width=1\linewidth]{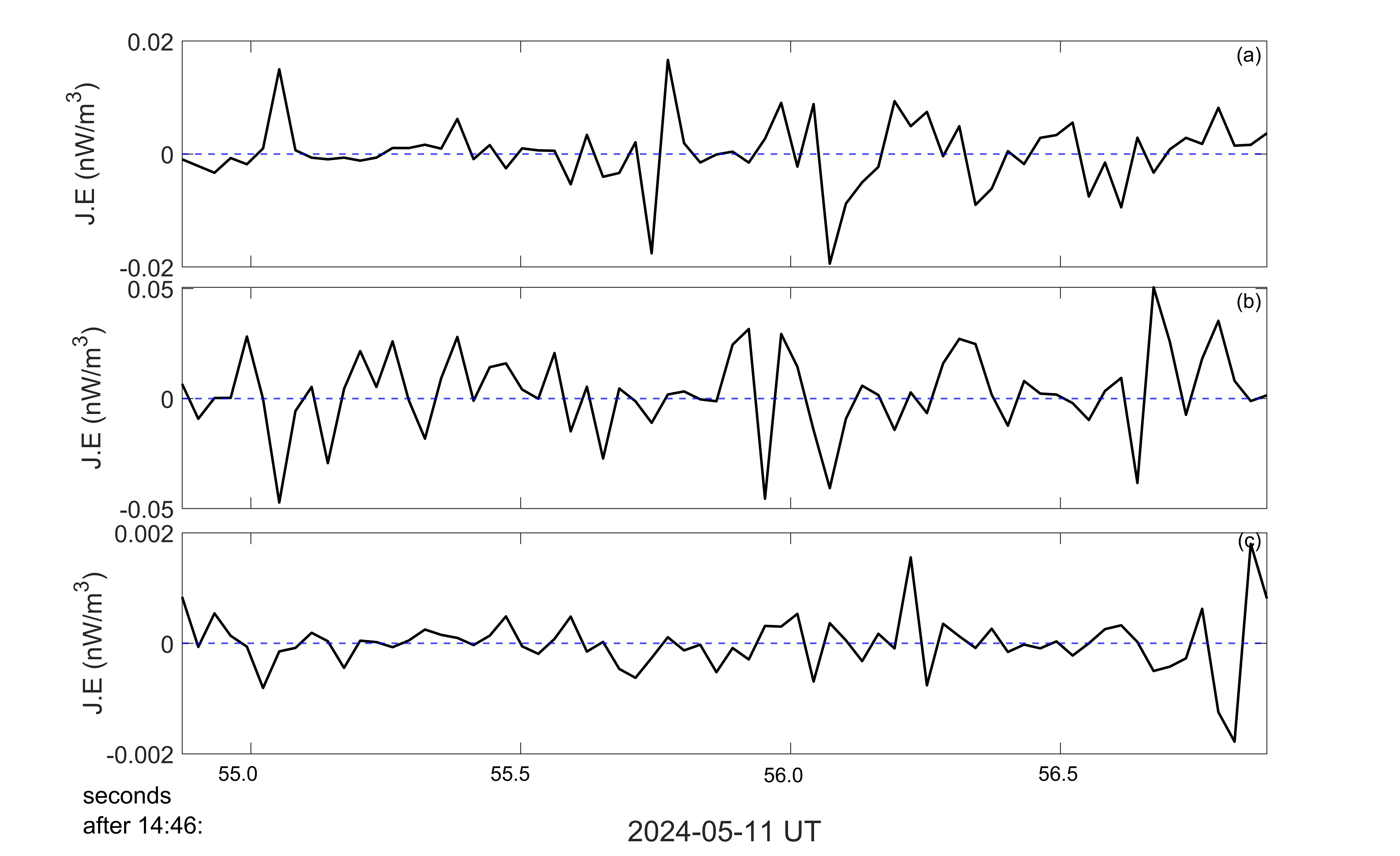}
\caption*{\textbf{Supplementary Figure 3}: Energy transfer parameter \( \vec{J} \cdot \vec{E} \) for: (a) Whistler wave, (b) Fundamental band of ECH wave, (c) Langmuir wave.}
\label{fig:SupplementaryFig3}
\end{figure}

\end{document}